\documentclass[11pt,a4paper]{article}

\usepackage{ijcks}
\usepackage{times}
\usepackage{latexsym}
\usepackage{url}
\usepackage[dvipdfm]{color,graphicx} 

\title{Rebels Lead to the Doctrine of the Mean: \\ Opinion Dynamic in a Heterogeneous DeGroot Model}
\author{Zhigang Cao\affilA \and Mingmin Yang\affilA \and Xinglong Qu\affilA\and Xiaoguang Yang\affilA\\
\affilA Key Laboratory of Management, Decision \& Information Systems,\\  Academy of Mathematics
and Systems Science,\\ Chinese Academy of Sciences,Beijing, 100190, P.R. China.\\
{\tt zhigangcao@amss.ac.cn}}

\summary{
We study an extension of the DeGroot model where part of the players may be rebels. The updating rule for rebels is quite different with that of normal players (which are referred to as conformists): at each step a rebel first takes the opposite value of the weighted average of her neighbors' opinions, i.e. 1 minus that average (the opinion space is assumed to be [0,1] as usual), and then updates her opinion by taking another weighted average between that value and her own opinion in the last round. We find that the effect of rebels is rather significant: as long as there is at least one rebel in every closed and strongly connected group, under very weak conditions, the opinion of each player in the whole society will eventually tend to 0.5.
}

\begin{document}
\maketitle


\section{Introduction}
Social learning is an old yet still booming research field, attracting more and more attention from economists, sociologists, physicists, and even computer scientists and game theorists (Jackson, 2009; Castellano, 2009; Lorenz, 2007). Based on whether players are fully rational or bounded rational, this field can be roughly divided into two branches, Bayesian learning and non-Bayesian learning (Jackson, 2009; Acemogluy and Ozdaglar, 2010). One common assumption in current research is that there is a learning topology, which is usually represented by a network (perhaps weighted and directed), indicating who learns from whom. One main difference between the research of Bayesian learning and that of non-Bayesian learning is that the learning in the first branch is repeated for each player, but usually one-shot in the second branch, because repeated Bayesian learning on networks is formidably difficult, as shown by Syngjoo et al. (2008). It's very hard to say which of the two learning rules is more realistic, because the former is so complicated that it is beyond the ability of ordinary human being, and conversely the latter is over simplified that people may not be so naive at all in the real world. However, for academic studies, they can both serve as very good benchmarks. In this paper, we shall concentrate on one particular model of the non-Bayesian learning, the DeGroot model (DeGroot, 1974).

In almost all learning models studied so far, it is assumed that people believe that the opinions (or information, beliefs) that their neighbors (or friends) hold are not only {\it valuable}, but also {\it correct} (to certain degree). To understand the difference between valuable and correct, just consider the situation where there is a liar who always lies about his real opinion by telling the opposite. Although the liar's opinion may not be correct at all, it's still valuable for his neighbors, because they can interpret this opinion by taking an opposite once again. And what's more, it's also usually assumed that the initial opinion that each person hold is indeed correct to certain degree. To be more specific, it is usually assumed that the (perhaps weighted) average opinion of the whole society (often required to be large enough) is (at least approximately) true. Consequently, for any given learning rule, whether people in the society can aggregate the scattered opinions into the ultimately true one, i.e. whether wise consensus (in various senses) is reachable,  is one of the core problems studied in the field of social learning.

The situation with naive liars can be easily transformed into classical models as shown by the above argument, as long as whether a player is a liar or not is known to all her neighbors. However, the other situation where there are players who don't believe that the opinions of their neighbors are correct at all, for various reasons,  can not be transformed into the classical models. In this paper, we shall study a special case of the above situation where there are {\it rebels}, i.e. players who always take the opposite opinions to their neighbors' overall opinions.

 We are not going to discuss whether rebels are rational or not, but their effect on the classical learning models. First of all, rebels do exist in the real world. They behave like this either because of their characters or because they believe this is fashionable. In fact, the term rebel comes from Jackson (2009), who formulates a game model called the fashion game (see also Cao and Yang, 2011). Another evidence comes from Krugman (1996), who argues that ``some intellectuals reject comparative advantage simply out of a desire to be intellectually fashionable", because ``in a culture that always prizes the avant-garde, attacking that icon is seen as a way to seem daring and unconventional".

Our research is based directly on the seminal DeGroot model (DeGroot, 1974), which is still one of the most basic models studied in non-Bayesian learning. In the DeGroot model, people update their opinions at each step simply by taking a weighted average of their neighbors' opinions. And the updatings of all players are done at each step simultaneously. Historically, this naive learning rule has long been thought of as too simple to lead to a wise consensus (Sobel, 2000). Interestingly, it is shown recently by Golub and Jackson (2010) that there exists a fairly broad collection of networks where wise consensus can be reached.

Our finds are quite surprising, the effect of rebels is not only remarkable but even dominant. Under very weak conditions, very few rebels can lead the whole society to the doctrine of the mean, i.e. all people (not only rebels) hold eventually an opinion of 0.5 (the opinion space as usually assumed is [0,1]), and this is regardless of initial opinions.

Our model can be taken as a heterogeneous model, i.e. there are more than one types of agents, and our study also echoes the question raised by Golub and Jackson (2010) that ``can a small admixture of different agents significantly change the group's behavior?" Our answer is definitely yes for the situation with rebels.

 As stated in the first sentence of this section, social learning is a typical multidisciplinary field. This paper follows the research thread driven by economists. For more knowledge about this thread, please refer Jackson (2009), Acemoglu and Ozdaglar (2010),  and Golub and Jackson (2010). In the thread driven by physicians, it is usually called opinion dynamics and rarely social learning (in comparison, in that driven by economists, the terms social learning and opinion dynamics are interchangeably used). In the study of opinion dynamics by physical scientists, the effect of rebels, where Galam calls them {\it contrarians}, has already been studied (Galam, 2004; Galam, 2008). We note that there are huge differences between the model of Galam and that of ours. To be specific, Galam studies a voter model, i.e. there are only two possible opinions, 0 and 1, while opinions in our model are continuous. And the interaction process used in his model (mainly random grouping at each step), is completely different from ours.  However, very interestingly, the main finding of Galam has a very similar spirit with ours: he finds that contrarians in the voter game tend to lead to hung elections (i.e. the election results are almost 50:50). A very actively studied model in opinion dynamics is the HK model, please refer Lorenz (2007) for more literature. For more models of opinion dynamics, please refer Galam (2008) and Castellano et al. (2009).

 The rest of this paper is organized as follows. Section \ref{description} gives a formal description of our model, as well as the necessary preliminaries that will be used in later sections. Section \ref{result} is the main body of this paper, where the theoretical results are provided. Section \ref{conclusion} concludes this paper with several further remarks.

  \section{Model Description and Preliminaries\label{description}}
  We are given a set of players $N=\{1,2,\cdots,n\}$. Initially, each player $j\in N$ holds an opinion $x_{j}(0)$. As usual, we assume that $x_{j}(0)\in [0,1]$. The learning topology is represented by a stochastic matrix $A$, i.e. each entry of $A$ is nonnegative, and each row of $A$ sums to 1. $A_{jk}$ is the weight that player $j$ places on $k$. The larger $A_{jk}$ is, the more important $k$ is in the eyes of $j$. To put it another way, $A_{jk}$ is the extent to which player $j$ believes that the opinion of $k$ can represent that of the whole society (herself excluded).  Thus, the value of $A_{jk}$ is private to $j$. In the opinion updating process, player $j$ considers the opinion of $k$ iff $A_{jk}>0$, in which case we say that $k$ is a neighbor of $j$. Notice that the relationship of being neighbors may not be symmetric, i.e. it may well happen that $k$ is $j$'s neighbor, but not the converse.

  Each player has a type of either a {\it conformist} or a {\it rebel}, which is pre-given and fixed.
  For all $j\in N$, let $x_j(t)$ be her opinion at time $t$. Rebels and conformists distinguish each other by their different opinion updating rules. The updating rule of a conformist is exactly the same as in the DeGroot model. If $j$ is a conformist, then at time $t+1$ she takes a weighted average over all the opinions at time $t$ (her own opinion at time $t$ will also be considered, as long as $A_{jj}>0$), i.e. $x_j(t+1)=\sum_{k\in N}A_{jk}x_{k}(t)$.

  If $j$ is a rebel,  then unlike a conformist who tries to hold an opinion that is as close as possible to the overall opinion of the whole society, she tries to be different with others, i.e. she tries first to detect the overall opinion of the others, and then take an opposite (1 minus that overall opinion, because we assume the opinion space is $[0, 1]$).
  The critical issue is how to deal with the opinion of her own in the last round. We assume that she desires to be consistent with herself.
  It's kind of absurd for her to treat her own opinion equally as the other opinions, because although she likes to hold opposite opinions with the others, she should not refute herself. Just picture the extreme case where a rebel $j$ puts very tiny weights on the others (perhaps because she does not get along with her neighbors, and thus is rather unsure about their real opinions), if her own opinion in the last round is treated the same as that of her neighbors, then in each round she will take an almost opposite opinion as to the last round. This is quite quirky, and  very few people, if any,  have this kind of bizarre personality.

  To distinguish the different treatment of a rebel $j$ on her own opinion, we use $\lambda_{j}\in [0,1]$ to denote her level of {\it confidence}, i.e. the weight she puts on herself. For convenience, we assume that $A_{jj}=0$ and $\sum_{j\in N}A_{jk}=1$, and put an overall weight of $1-\lambda_j$ on the average opinion of her neighbors. The updating rule of a rebel $j$ is $x_{j}(t+1)=\lambda_{j} x_{j}(t)+(1-\lambda_{j})(1-\sum_{k\in N}A_{jk}x_{j}(t))$.

  To get a uniform updating formula for both rebels and conformists, for each conformist $j$ we also use $\lambda_j$ to denote her level of confidence and assume that $A_{jj}=0$. To sum up, we have
  {\tiny \begin{equation}x_j(t+1)=\left\{\begin{array}{ll}\lambda_jx_j(t)+(1-\lambda_j)\sum_{k\in N}A_{jk}x_{k}(t)&if~u_j=1\\
  \lambda_{j} x_{j}(t)+(1-\lambda_{j})\left(1-\sum_{k\in N}A_{jk}x_{j}(t)\right)&if~u_j=0\end{array}\right..\label{up1}\end{equation}}

  Notice that $A$ is still a stochastic matrix and recall our assumption that \begin{equation}A_{jj}=0,\forall j\in N.\end{equation}

  For technical reasons, we further assume in this paper that all the confidence levels are identical, and we use a new symbol $\lambda$ to denote this value, i.e.
  \begin{equation}\lambda_j=\lambda,\forall j\in N.\label{a2}\end{equation}

In the rest of this section, we shall provide several necessary concepts and preliminaries. Associated with each stochastic matrix $A$ is a digraph $\mathcal{G}(A)$: the nodes are naturally $N=\{1,2,\cdots,n\}$, and there is an edge going from $j$ to $k$ if and only if $A_{jk}>0$. To study the structure of $\mathcal{G}(A)$ is usually more intuitive than to study $A$ directly, and tools of graph theory can also be conveniently applied.
$\mathcal{G}(A)$ is called {\it strongly connected} if and only if for each pair of nodes $(j,k)$, there is a sequence of directed edges leading from $j$ to $k$. It turns out that $\mathcal{G}(A)$ is  strongly connected  if and only if $A$ is {\it irreducible}. We present this fact formally by the following definition and lemma.

{\bf Definition 1.} {\it [Meyer, 2000, p671] $A_{n\times n}$ is said to be a reducible matrix when there exists a permutation
matrix $P$ such that $P^TAP=\left(\begin{array}{cc}X&Y\\
0&Z\end{array}\right),$
where $X$ and $Z$ are both square. Otherwise, it is called irreducible.}

{\bf Lemma 1.} {\it [Meyer, 2000, p671] $A$ is an irreducible matrix if and only if $\mathcal{G}(A)$ is strongly connected.}

In this paper, we shall use the term strongly connected and irreducible interchangeably.

Given a stochastic matrix $A$ and the associated digraph $\mathcal{G}(A)$, a sequence of nodes $(j_1,j_2,\cdots,j_s)$ is called a {\it cycle} if and only if (i) $A_{j_lj_{l+1}}>0$ for all $1\leq l\leq s$, where $j_{s+1}\equiv 1$; (ii) no node is repeated.  We note that this concept is called {\it simple cycle} in Golub and Jackson (2010).

Another important concept is {\it aperiodic}, which describes a digraph that the greatest common divisor of the lengths of its cycles is 1. It turns out that for strongly connected graphs, the concept of aperiodic is equivalent to a basic concept {\it primitive} in matrix theory. Following the notations of Meyer (2000), we shall use $\rho(A)$ throughout the rest of this paper to denote the spectrum radius of matrix $A$, i.e. the largest absolute value of all the eigenvalues of $A$ (which might be complex). Also, we use $\sigma(A)$ to denote the set of all the eigenvalues of $A$, then\begin{equation}\rho(A)=\max\{|r|: r\in \sigma(A)\}.\end{equation}

The exact definition of primitive is as follows.

{\bf Definition 2.} {\it [Meyer, 2000, p674] A nonnegative irreducible matrix $A$ having only one eigenvalue,
$r=\rho(A)$, on its spectral circle is said to be a primitive matrix.}

Notice that the definition used in Golub and Jackson (2010) is not the same as the above one. In fact, their definition is a characterization of primitive matrices. The following formal equivalence relationship between aperiodic and primitive is first proved by Perkins (1961), and the form we present is directly from  Golub and Jackson (2010).

{\bf Lemma 2.} {\it Assume $A$ is stochastic and the associated digraph $\mathcal{G}(A)$ is strongly connected. Then $\mathcal{G}(A)$ is aperiodic if and
only if $A$ is primitive.}

To finish this section, we present one more basic result of matrix theory. Other deeper results used in the next section will be provided there.

{\bf Lemma 3.} {\it [Meyer, 2000, p618] For $A\in \mathcal{C}^{n\times n}$, where $\mathcal{C}$ is the set of complex numbers, the following statements are equivalent.

(i) The Neumann series $I+A+A^2+\cdots$  converges.

(ii) $\rho(A)<1$.

(iii) $\lim_{k\rightarrow\infty}A^k=0$.

In which case, $(I-A)^{-1}$ exists and $\sum_{k=0}^{\infty}A^k=(I-A)^{-1}$.}
\section{Main Results\label{result}}
We study first the easier case where all players are rebels, and then the general case with partial rebels and partial conformists.
\subsection{The special case with all rebels}

When all players are rebels, the updating rule can be represented conveniently in vector and matrix notations. In fact, let $x(t)=(x_1(t),x_2(t),\cdots,x_n(t))^T$ and ${\bf 1}=(1,1,\cdots,1)^T$, the following updating rule is obvious due to (\ref{up1}) and (\ref{a2}).

\begin{equation}x(t+1)=\lambda x(t)+(1-\lambda)({\bf 1}-Ax(t)).\label{dm}
\end{equation}

The first property we study about (\ref{dm}) is convergence, by which we mean that $\lim_{t\rightarrow \infty}x_j(t)$ exists for all $j\in N$, regardless of initial values $x(1)$.

{\bf Theorem 1.} {\it In the special case of the heterogeneous DeGroot model with all rebels, suppose the learning topology $A$ is stochastic and the associated digraph $\mathcal{G}(A)$ is strongly connected.

(i) When $\lambda=0$, dynamic (\ref{dm}) is divergent.

(ii) When $\lambda\neq 0$, if $-1\notin \sigma(A)$, then dynamic (\ref{dm})  converges to $\frac{1}{2}{\bf 1}$, regardless of the initial opinions $x(1)$.}

{\bf Proof.} Dynamic (\ref{dm}) can be rewritten as $$x(t+1)=(\lambda I-(1-\lambda) A)x(t)+(1-\lambda){\bf 1}.$$

Let $B=\lambda I-(1-\lambda) A$, we have {\scriptsize $$x(t+1)=B^tx(1)+(1-\lambda)(I+B+B^2+\cdots+B^{t-1}){\bf 1}.$$}

It's valuable to notice that the convergence of $x(t)$ might not be exactly the same as the convergence of the Neumann series $I+B+B^2+\cdots+B^{t-1}$. In fact, the latter might well be strictly stronger, because here ${\bf 1}$ is not an arbitrary vector but fixed.

The fact that ${\bf 1}$ is an eigenvector of any stochastic matrix makes our analysis quite easy.  In fact, since $A$ is a stochastic matrix, it's easy to check that $B^t{\bf 1}=(2\lambda-1)^t{\bf 1}$, and thus $x(t+1)=B^tx(1)+\frac{1}{2}(1-(2\lambda-1)^{t}){\bf 1}$.

(i) When $\lambda=0$, dynamic (\ref{dm}) does not converge for $x(1)={\bf 1}$, and thus is divergent. In fact, it can be observed in this case that $x(t+1)=\frac{1-(-1)^{t+1}}{2}$.

(ii) When $\lambda\neq 0$ and $-1\notin \sigma(A)$, we know by definition of $B$ that $\rho(B)<1$. In fact, for each $r\in \sigma(B)$ (notice that $r$ might be complex), there exists a complex number $xi+y\in \sigma(A)$, $x^2+y^2\leq 1$,  such that $$r=\lambda-(1-\lambda)(xi+y).$$ Therefore \begin{eqnarray*}|r|^2&=&(\lambda-(1-\lambda)y)^2+((1-\lambda)x)^2\\
&=&\lambda^2+(1-\lambda)^2(x^2+y^2)-2\lambda(1-\lambda)y\\
&\leq&\lambda^2+(1-\lambda)^2+2\lambda(1-\lambda)\\
&=&1,\end{eqnarray*}
and equality holds if and only if $y=-1$, i.e. $-1\in \sigma(A)$.

Lemma 3 tells us that $\lim_{t\rightarrow \infty}B^t=0$, and thus $\lim_{t\rightarrow \infty}B^tx(1)=0$ for all $x_1$. On the other hand, $|2\lambda-1|<1$ implies that $\lim_{t\rightarrow \infty}(2\lambda-1)^{t}=0$. Therefore dynamic (\ref{dm}) converges to $\frac{1}{2}{\bf 1}$ regardless of $x(1)$, and hence the theorem.
$\Box$

{\bf Corollary 1.} {\it In the special case of the heterogeneous DeGroot model with all rebels, suppose the learning topology $A$ is stochastic and the associated digraph $\mathcal{G}(A)$ is strongly connected. If $\lambda\neq 0$, $\mathcal{G}(A)$ is aperiodic, then dynamic (\ref{dm}) converges to $\frac{1}{2}{\bf 1}$, regardless of $x(1)$.}

{\bf Proof.} By Lemma 2 we know that  $A$ is primitive, and therefore 1 is the only eigenvalue that is on the spectral circle. Consequently, it is impossible for -1 to be an eigenvalue of $A$. By part (i) of Theorem 1, this corollary is valid. $\Box$

We are going to show next that when $\lambda\neq 0$, then $\mathcal{G}(A)$ must be very special to have (\ref{dm}) to be divergent. And consequently in this case, (\ref{dm}) is convergent for a very broad classes of learning topologies. To demonstrate this result, we need to define a special class of digraphs, which is a slight generalization of bipartite graphs.

{\bf Definition 3.} {\it [Brualdi and Cvetkovic, 2009, p176] Suppose $G$ is a digraph with node sets $N$. If $N$ can
be partitioned into $h$ nonempty sets $N_0, N_1,\cdots,N_h$,  such that each
edge of $G$ has its initial node in some $N_l$ and its terminal node
in $N_{l+1}$ (subscripts considered modulo $h$), then $G$ is called cyclically $h$-partite. }

{\bf Corollary 2.} {\it In the special case of the heterogeneous DeGroot model with all rebels, suppose the learning topology $A$ is stochastic and the associated digraph $\mathcal{G}(A)$ is strongly connected. If $\lambda\neq 0$, and dynamic (\ref{dm}) does not converge, then $\mathcal{G}(A)$ is a cyclically $h$-partite graph for some $h\geq 2$.}

{\bf Proof.} By part (ii) of Theorem 1 we know that $A$ is imprimitive ($\{-1,1\}\subseteq \sigma(A)$). Brualdi and Cvetkovic (2009, p176) tell us that for each strongly connected imprimitive stochastic matrix, its associated digraph is $h$-partite.  Hence the corollary. $\Box$

Brualdi and Cvetkovic (2009, p176) tell us more. In fact, the parameter $h$ in Corollary 2 is exactly the {\it index of imprimitivity}, i.e. the number of eigenvalues on the spectral circle.

We finish this subsection by noting that whether a stochastic matrix has an eigenvalue of -1, a property that  is very crucial to the convergence of (\ref{dm}) as shown by Theorem 1, can be checked very efficiently. In fact, $-1\in \sigma(A)$ if and only if the determinant of $I+A$ is zero.

\subsection{The general case with partial rebels}
Recall that $\forall j\in N$, $u_j=1$ means that she is a conformist, and $u_j=0$ a rebel. Let $U=diag(u_1,u_2,\cdots,u_n)$, i.e. the $n\times n$ matrix whose diagonal entries are $u_1,u_2,\cdots,u_n$ and all the other entries are zero. Then the updating rule is {\scriptsize \begin{equation}x(t+1)=\lambda x(t)+(1-\lambda)(UAx(t)+(I-U)({\bf 1}-Ax(t))).\label{dm2}\end{equation}}

{\bf Theorem 2.} {\it In the heterogeneous DeGroot model with partial rebels, suppose the learning topology $A$ is stochastic and the associated digraph $\mathcal{G}(A)$ is strongly connected.

(i) When $\lambda=0$, if $\rho((2U-I)A)\neq 1$, then dynamic (\ref{dm2}) converges to $\frac{1}{2}{\bf 1}$.

(ii) When $\lambda\neq 0$, if $1\notin \sigma((2U-I)A)$, then dynamic (\ref{dm2}) converges to $\frac{1}{2}{\bf 1}$.}

{\bf Proof.}  Dynamic (\ref{dm2}) can be rewritten as {\scriptsize $$x(t+1)=(\lambda I+(1-\lambda)(2U-I)A)x(t)+(1-\lambda)(I-U){\bf 1}.$$}

Let $B=\lambda I+(1-\lambda)(2U-I)A$, we have {\scriptsize $$x(t+1)=B^tx(1)+(I+B+B^2+\cdots+B^{t-1})(I-U){\bf 1}.$$}

Since $|(2U-I)A|\leq A$, i.e. the absolute value of each entry of $(2U-I)A$ is no more than the corresponding value of $A$, we know from Lemma 4 below that \begin{equation}\rho((2U-I)A)\leq \rho(A)=1.\end{equation}

(i) When $\lambda=0$, $B=(2U-I)A$. If $\rho((2U-I)A)\neq 1$, then $\rho(B)<1$, and thus by Lemma 3 we know that $\lim_{t\rightarrow \infty}B^t=0$, and the Neumann series $I+B+B^2+\cdots+B^{t-1}$ converges to $(I-B)^{-1}=(I-(2U-I)A)^{-1}$. Therefore, $(I+B+B^2+\cdots+B^{t-1})(I-U){\bf 1}$ converges to $(I-(2U-I)A)^{-1}(I-U){\bf 1}$.

Because
\begin{eqnarray*}(I-(2U-I)A){\bf 1}&=&{\bf 1}-(2U-I){\bf 1}\\
&=&2(I-U){\bf 1},\end{eqnarray*}

 we get $$(I-(2U-I)A)^{-1}(I-U){\bf 1}=\frac{1}{2}{\bf 1}.$$

(ii) When $\lambda\neq 0$, using the same argument as in the proof to part (ii) of Theorem 1, we know that $\rho(B)=1$ is equivalent to $$1\in \sigma((2U-I)A).$$

 Hence the theorem. $\Box$

{\bf Definition 3.} {\it Suppose $G$ is a digraph, and each node belongs to one of the two types, conformists and rebels. We say $G$ is a rebel-bipartite graph if and only if there is no cycle with an odd number of rebels.}

To prove the next theorem, we need the following standard result from matrix theory.

{\bf Lemma 4.} {\it [Wielandt's Theorem, Meyer, 2000, p675] If $|B|\leq A_{n\times n}$, where $A$ is irreducible, then $\rho(B)\leq\rho(A)$. If equality
holds (i.e., if $\mu=\rho(A)e^{i\phi}\in\sigma(B)$ for some $\phi$), then
$B=e^{i\phi}D^{-1}AD$  for some$$D=\left(\begin{array}{cccc}e^{i\theta_1}&~&~&~\\
~&e^{i\theta_2}&~&~\\
~&~&\ddots&~\\
~&~&~&e^{i\theta_1}\\
\end{array}\right),
$$
and conversely.}

{\bf Theorem 3.} {\it In the heterogeneous DeGroot model with partial rebels, suppose the learning topology $A$ is stochastic and the associated bigraph $\mathcal{G}(A)$ is strongly connected. When $\lambda\neq 0$, if dynamic (\ref{dm2}) does not converge, then $\mathcal{G}(A)$ is a rebel-bipartite graph.}

{\bf Proof.} By Theorem 2, the hypothesis that dynamic (\ref{dm2}) does not converge means that $$\rho((2U-I)A)=1.$$

 By Lemma 4, there exits $\theta_1,\cdots,\theta_n\in[0,2\pi)$ such that
\begin{equation}(2U-I)A=D^{-1}AD,\label{e1}\end{equation} where $D=\left(\begin{array}{cccc}e^{i\theta_1}&~&~&~\\
~&e^{i\theta_2}&~&~\\
~&~&\ddots&~\\
~&~&~&e^{i\theta_n}\\
\end{array}\right).
$

Equality (\ref{e1}) says that
\begin{equation}A_{jk}=e^{-i\theta_j}e^{i\theta_k}A_{jk},~\mbox{if}~j~\mbox{is~a~conformist},\label{e2}\end{equation}
and \begin{equation}-A_{jk}=e^{-i\theta_j}e^{i\theta_k}A_{jk},~\mbox{if}~j~\mbox{is~a~rebel}.\label{e3}\end{equation}

For each $A_{jk}>0$, equations (\ref{e2})(\ref{e3}) tell us that
\begin{equation}\theta_j=\theta_k, ~\mbox{if}~j~\mbox{is~a~conformist},\label{e4}\end{equation}
\begin{equation}|\theta_j-\theta_k|=\pi, ~\mbox{if}~j~\mbox{is~a~rebel}.\label{e5}\end{equation}

With the above discussions in hand, we are now ready to prove that $\mathcal{G}(A)$ is a rebel-bipartite graph. Suppose on the contrary that there exists a directed graph with an odd number of rebels. Let $j_1,j_2,\cdots,j_{2k-1}$ be all the rebels on such a cycle, and this is the order that they are allocated, (i.e. $j_2$ is the first rebel that we meet if we start from $j_1$ and walk along the cycle in the direction that is consistent with the graph, and so on). By (\ref{e4})(\ref{e5}) we know that $$|\theta_{j_1}-\theta_{j_2}|=\pi,\cdots,|\theta_{j_{2k-2}}-\theta_{j_{2k-1}}|=\pi.$$

W.l.o.g., suppose $\theta_{j_1}-\theta_{j_2}=\pi$, i.e.  $\theta_{j_1}=\theta_{j_2}+\pi$. Since $\theta_{j_1}\in[0,2\pi)$, this can only happen when $\theta_{j_2}\in[0,\pi)$. And therefore
$|\theta_{j_2}-\theta_{j_3}|=\pi$ can only happen when $\theta_{j_3}=\theta_{j_2}+\pi$, because if $\theta_{j_2}=\theta_{j_3}+\pi$ we would get $\theta_{j_3}<0$, which is impossible by hypothesis. Hence $$\theta_{j_1}=\theta_{j_3}.$$

Repeating the above argument, we will get $\theta_{j_1}=\theta_{j_3}=\cdots=\theta_{j_{2k-1}}=\theta_{j_2}=\cdots=\theta_{j_{2k-2}}$, i.e. all the $\theta$'s are identical, which is impossible. Hence the theorem. $\Box$
\subsection{Further discussions}

All the discussions in the preceding subsections assume that the learning topology is strongly connected. If we dump this assumption, then similar results still hold. In fact, as explored by Golub and Jackson (2010), each closed and strongly connected group of the whole society evolves completely like a small strongly connected society, because their opinions are not affected by players outside of this group at all. It's interesting to investigate behaviors of the players that are not in any closed and strongly connected group.

As to convergence rate, we know from standard matrix theory that it is determined in the special case with all rebels by the second largest eigenvalue of $\lambda I-(1-\lambda)A$, which is $\lambda-(1-\lambda)r$, where $r$ is the second smallest eigenvalue of $A$. It might be much more complicated for the general case, because the eigenvalues of $\lambda I+(1-\lambda)(2U-I)A$ might have no connection with those of $A$ at all.

\section{Conclusions\label{conclusion}}
We study a heterogeneous DeGroot model in this paper. Analysis shows that the effect of rebels is significant: under very weak conditions they will always lead the society to doctrine of the mean. This result is more or less surprising, because at first sight the rebels seem to be really radical. Our result confirms further the mediation role of rebels, which is first discovered by Galam (2004), i.e. they tend to make things more equal. Further directions include giving sufficient and necessary conditions for convergence and analyzing the more realistic situation where different players may have different confidence levels. It's also very interesting to investigate the effect of the other kind of rebels who go to extremes. To be precise, if the overall opinion of her neighbors is 0.3, then she will hold an opinion of 1, and in the case that her neighbors hold an opinion of 0.7, she will choose 0. They might be rebels in the real sense.


\begin{thebibliography}{00}
\bibitem{ao10}D.Acemogluy and A. Ozdaglar. Opinion Dynamics and Learning in Social Networks. Dynamic Gams and Applications, 1(1): 3-49, 2010.
\bibitem{bc09}R.A. Brualdi and D. Cvetkovic. A combinatorial approach to matrix theory and its applications. Taylor and Francis Group, LLC, Boca Raton, 2009.
\bibitem{cy11}Z.Cao and X.Yang. The fashion game and the fashion curse. Available at SSRN:
http://ssrn.com/abstract=1767863.
\bibitem{cfl09}C.Castellano, S.Fortunato, W. Loreto. Statistical physics of social dynamics. Review of Modern Physics, 81: 591-646, 2009.
\bibitem{d74}MH. DeGroot. Reaching a Consensus, Journal of the American Statistical
Association, 69: 118-121, 1974.
\bibitem{m00}CD. Meyer. Matrix Analysis and Applied Linear Algebra. SIAM, Philadelphia, 2000.
\bibitem{dvz03}P. DeMarzo, D. Vayanos, and J. Zwiebel. Persuasion bias, social influence, and unidimensional
opinions. Quarterly Journal of Economics, 118: 909-968, 2003.

\bibitem{j09}M. O. Jackson. Social and Economic Networks. Princeton University Press, Princeton, 2009.
\bibitem{g04}S. Galam. Contrarian deterministic effects on opinion dynamics:``the hung elections scenario". Physica A: Statistical and Theoretical Physics. 333: 453-460, 2004.
    \bibitem{g08}S. Galam. Social Physics: A Revew of Galam Models. International Journal of Modern Physics C. 19(3), 2008: 409-440.
\bibitem{gj10}B. Golub and M.O. Jackson. Naive Learning in Social Networks and the
Wisdom of Crowds. American Economic Journal: Microeconomics, 2(1): 112-149,  2010.
\bibitem{k96}P. Krugman.  Ricardo's difficult idea. Paper for Manchester conference on free trade, March 1996. (available on his official web page)
    \bibitem{l07} J. Lorenz. Continuous Opinion Dynamics under Bounded Confidence: A Survey. International Journal of Modern Physics, 18(12): 1819-1838, 2007.
        \bibitem{p61}P. Perkins. A Theorem on Regular Matrices. Pacific Journal of Mathematics, 11(4): 1529-33, 1961.
    \bibitem{s00}J. Sobel. Economists¡¯ Models of Learning. Journal of Economic Theory, 94(2): 241-261, 2000.
    \bibitem{sgk08}C. Syngjoo, D. Gale, and S. Kariv. Sequential Equilibrium in Monotone
Games: A Theory-Based Analysis of Experimental Data. Journal of Economic Theory, 143(1):
302-30, 2008.
\end{thebibliography}
\end{document}